\documentclass[preprint,12pt,3p]{elsarticle}

\usepackage{amsmath,amssymb,amsfonts}
\usepackage{mathrsfs}
\usepackage{chemformula}
\usepackage{color}
\usepackage[normalem]{ulem}
\usepackage{stmaryrd}
\usepackage{diagbox}
\usepackage[parfill]{parskip}

 \biboptions{sort&compress}

\newcommand{\blue}[1]{\textcolor{black}{#1}}

\newcommand{\beq}{\begin{equation}}
\newcommand{\eeq}{\end{equation}}

\newcommand\Pec{\mbox{\textit{Pe}}}

\newcommand\Lew{\mbox{\textit{Le}}}

\journal{Progress in Scale Modeling, an International Journal}

\begin{document}

\begin{frontmatter}

\title{Tricritical point as a crossover between type-I$_s$ and type-II$_s$ bifurcations}

\author{Prabakaran Rajamanickam, Joel Daou}
\address{Department of Mathematics, University of Manchester, Manchester M13 9PL, UK}

\begin{abstract}
 A tricritical point as a crossover between (stationary finite-wavelength) type-I$_s$ and (stationary longwave) type-II$_s$ bifurcations is identified in the study of diffusive-thermal (Turing) instability of flames propagating in a Hele-Shaw channel in a direction transverse to a shear flow. Three regimes exhibiting different scaling laws are identified in the neighbourhood of the tricritical point. For these three regimes, sixth-order partial differential equations are obtained governing the weakly nonlinear evolution of unstable solutions near the onset of instability.  These sixth-order PDES may be regarded as the substitute for the classical fourth-order Kuramoto--Sivashinsky equation which is not applicable  near the tricritical point.
\end{abstract}

\begin{keyword}
    Tricritical point \sep Sixth-order PDEs \sep Kuramoto–Sivashinsky equation \sep Weakly nonlinear analysis
\end{keyword}

\end{frontmatter}

\section{Tricritical point}

The concept of tricritical point is a well recognized topic in the theories of phase transition. It is a point in the phase diagram where the curves of first-order and second-order phase transitions meet~\cite{landau2013statistical}. On the other hand, these two phase transitions are qualitatively analogous to a stationary finite-wavelength (known as type-I$_s$) and a stationary longwave (known as type-II$_s$) bifurcations,  studied in the linear stability theory. The notation type-I$_s$ and type-II$_s$ follow the terminology of the review paper~\cite{cross1993pattern} and the book~\cite[pp. 75-81]{cross2009pattern} with the subscript $s$ indicating a stationary or non-oscillatory bifurcation from a stable to an unstable state. Analogous to the tricritical point of the phase transition aforementioned, a tricritical point as a crossover between type-I$_s$ and type-II$_s$ bifurcations appears not to have been studied yet in the literature. In fact, we have encountered such a point in a recent investigation~\cite{daou2023diffusive} of the diffusive-thermal (Turing) instability of a premixed flame propagating in a Hele-Shaw channel in a direction transverse to the shear-flow direction. The flame stability in this case is governed by a dispersion relation, connecting the complex growth rate $\sigma$ of a normal-mode perturbation with its real wavenumber $k$, given by
\begin{equation}
  2\Gamma^2(1-\Gamma) + l(1-\Gamma+2\sigma+4\lambda k^2)=0 \quad \text{with} \quad \Gamma = \sqrt{1+4\sigma+4k^2}. \label{disp}
\end{equation}
Here, $l$ and $\lambda$ are two independent parameters, \blue{defined by}
\begin{equation}
    \blue{l\equiv \beta(\Lew-1) \in(-\infty,\infty), \qquad \qquad  \lambda \equiv \frac{\gamma\Pec^2}{1+\gamma\Pec^2} \in[0,1]}
\end{equation}
\blue{where $\beta$ and $\Lew$ are the Zeldovich and Lewis numbers and $\Pec$ the Peclet number based on the channel half-width and the flow (maximum) amplitude. As for $\gamma$, it is the Taylor-dispersion coefficient which is fully determined by the flow profile, equal to $8/945$ in the case of a plane Poiseuille flow and $1/20$ in the case of a Couette flow, for instance.} The dispersion relation yields, in general, three roots for $\sigma=\sigma(k)$, of which one of the roots is always real and satisfies the condition $\sigma(0)=0$. This particular root is the root that is involved in the type-I$_s$ and type-II$_s$ bifurcation, \blue{as it will be shortly confirmed from the implications of the Taylor expansion of $\sigma(k)$ which follows.} Henceforth we shall regard $\sigma$ to be real and refer only to this particular root. 

\begin{figure}[h!]
\centering
\vspace{-0.4cm}
\includegraphics[scale=0.6]{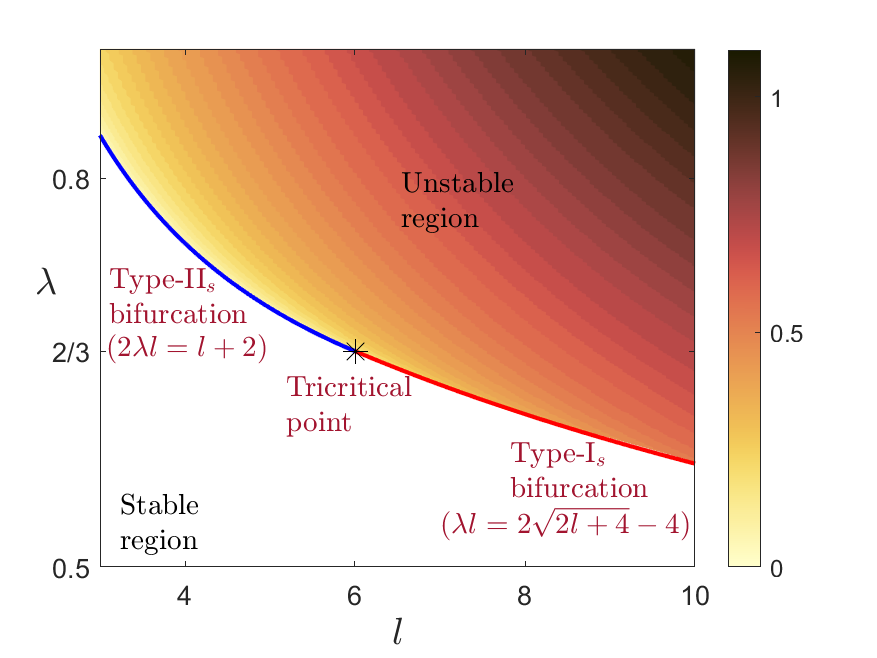}
\includegraphics[scale=0.6]{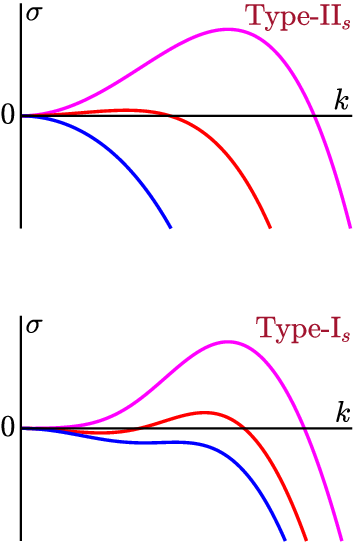}
\vspace{-0.4cm}
\caption{Left: Stability diagram in $l$-$\lambda$ plane. The colour scale indicates the wavenumber $k_m$ corresponding to the maximum growth rate; on the type-II$_s$ boundary, $k_m=k_c=0$ and on the type-I$_s$ boundary, $k_m=k_c\neq 0$, where $k_c$ is the critical wavenumber, i.e., $k_m$ at the onset of instability. \blue{The explicit formulas for type-I$_s$  and type-II$_s$ boundaries in the figure are derived in~\cite{daou2023diffusive} from the dispersion relation~\eqref{disp}; specifically, the type-II$_s$ boundary is derived from the condition $d^2\sigma/dk^2$ at $k=0$, whereas the type-I$_s$ boundary is derived from the conditions $\sigma=d\sigma/dk=0$ at $k\neq 0$.} Right: Schematic \blue{pictures} of type-I$_s$ and type-II$_s$ bifurcation. \blue{The blue lines represent dispersion curves just below the instability onset (i.e., for points lying in the white region in the left figure), whereas the other two lines represent dispersion curves just above the instability onset.}} 
\label{fig:stability}
\end{figure}

 A Taylor series expansion of $\sigma(k)$ at $k=0$ is given by
\begin{equation}
    \sigma = a k^2 + bk^4 + ck^6 + \cdots  \label{taylor}
\end{equation}
where 
\begin{equation}
    a = \frac{l}{2}(2\lambda-1)  -1, \quad b = \frac{l^2}{8}(2\lambda-1)^2 (l-6), \quad c = \frac{l^3}{16}(2\lambda-1)^3 (l-4)(l-10).  \label{taylorcoeff}
\end{equation}
The type-II$_s$ bifurcation is explained using only the first two terms of the Taylor series, provided $b<0$. Then the onset of instability corresponds to the condition $a=0$ or, equivalently the condition $d^2\sigma/dk^2|_{k=0}=0$ (see the schematic illustration in Fig.~\ref{fig:stability}).  The condition $b<0$ is needed since if $b>0$, then the critical wavenumber $k_c$ at the instability onset is necessarily non-zero and therefore  does not correspond to a type-II$_s$ bifurcation. It is evident from the definition of $b$ that as $l\to 6$, $b\to 0$; \blue{the vanishing of $b$ due to $l(2\lambda-1)=0$ is irrelevant since this corresponds to points lying elsewhere in the $l$-$\lambda$ plane than the region of interest under consideration.} This then defines a \textit{tricritical point} in the $l$-$\lambda$ plane (see Fig.~\ref{fig:stability}),
\begin{equation}
    (l,\lambda) = (6,2/3)
\end{equation}
as a point for which both $a=b=0$~\cite{landau2013statistical}. To characterise the neighbourhood of the tricritical point, we need the three-term expansion of the Taylor series with the condition $c<0$; $c=-4$ at the tricritical point. For $l>6$ (i.e., $b>0$), we have type-I$_s$ bifurcation which involves the change of sign of $\sigma$ at a critical wavenumber $k_c\neq 0$ where $d\sigma/dk=0$.

\section{The three regimes in the neighbourhood of the tricritical point}

The neighbourhood of the tricritical point can be examined by introducing two independent small parameters 
\begin{equation}
    \varepsilon = \frac{\lambda-2/3}{2/3}\ll 1, \qquad  \qquad \mu = \frac{l-6}{6} \ll 1. \label{epmu}
\end{equation}
The formula for $b$~\eqref{taylorcoeff} indicates that $b\simeq 3 \mu$ to leading order. The dispersion relation~\eqref{taylor} can therefore be written, in terms of $\mu$ and $\varepsilon$, as
\begin{equation}
    \sigma= ak^2 + 3\mu k^4 -4 k^6, \qquad a = 4\varepsilon + \mu +4\varepsilon\mu.  \label{disptri}
\end{equation}
The characteristic scale for $k$ and consequently that for $\sigma$ depend essentially on how $a$ or, equivalently how $\varepsilon$ compares with $\mu$.

\begin{figure}[h!]
\centering
\vspace{-0.4cm}
\includegraphics[scale=0.7]{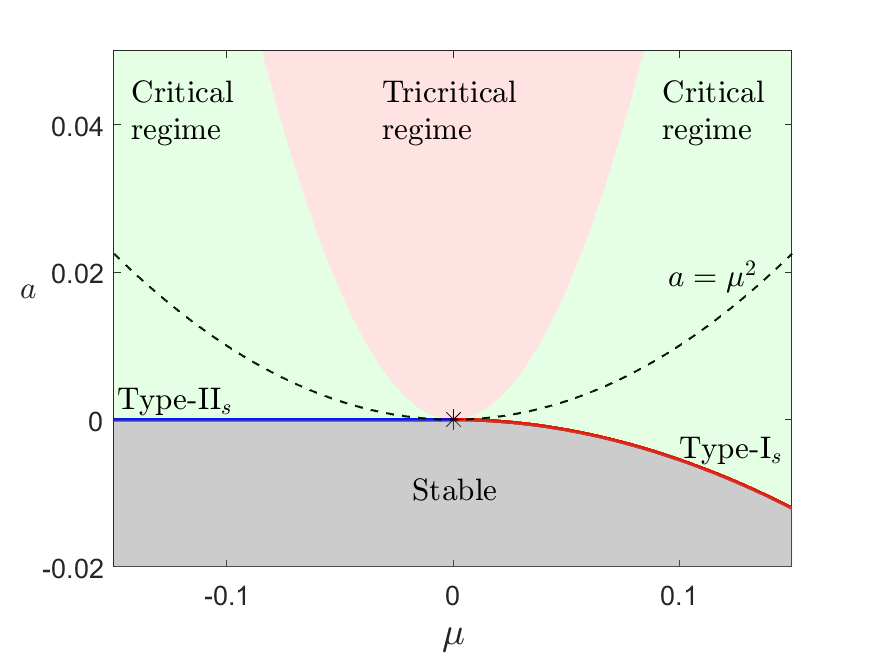}
\vspace{-0.4cm}
\caption{The neighbourhood of the tricritical point (origin) in the $\mu$-$a$ plane. \blue{The type-I$_s$ and type-II$_s$ boundaries are exactly the same as those in the left plot of Fig.~\ref{fig:stability}, which are given by $a=0$ for type-II$_s$ and $a=-(3\mu/4)^2 + \frac{1}{2}(3\mu/4)^3+ \cdots$ for type-I$_s$ boundary.}  The figure illustrates the two critical regimes on either side of the tricritical point, that flank a tricritical regime in the middle.} 
\label{fig:tri}
\end{figure}

It is instructive to visualize the neighbourhood of the tricritical point in the $\mu$-$a$ plane, as shown in Fig.~\ref{fig:tri}. Here the gray shaded region represents stable states and the remaining regions represent unstable states. The type-II$_s$ bifurcation, characterized by the condition $a=0$ is applicable for $\mu<0$, whereas the type-I$_s$ bifurcation is applicable for $\mu>0$. The parabola $a=\mu^2$ is plotted as a dashed curve for the purpose of reference. This figure identifies two critical regimes and a tricritical regime in the middle.

\subsection{Two critical regimes: $a= O(\mu^2)$}
First, let us consider the case in which $a= O(\mu^2)$. According to the definition of $a$~\eqref{disptri}, such a balance is possible only if
\begin{equation}
    4\varepsilon = -\mu  + (s+1) \mu^2 +\cdots,
\end{equation}
where $s$ is an arbitrary, order-unity constant. Thus, in critical regimes, $a$ is simply equal to $a=s\mu^2$ in the first approximation. The dispersion relation~\eqref{disptri} then simplifies to
\begin{equation}
    \sigma = s\mu^2k^2 + 3\mu k^4 - 4k^6. \label{triA}
\end{equation}
The dependence of $\sigma$ on $\varepsilon$ appears through the parameter $s$. The type-II$_s$ bifurcation for $\mu<0$ then corresponds to $s=0$ and the type-I$_s$ bifurcation for $\mu>0$ to $s=-9/16$. The latter result follows from the marginal condition $\sigma=0$ satisfied at $k_c^2=\mu(1+\sqrt{1+4s/3})/4$  where $d\sigma/dk=0$.

\subsection{Tricritical regime: $a\gg O(\mu^2)$}
\label{sec:tricriticalregime}
When $a\gg O(\mu^2)$, i.e., for points lying far above the parabola $a=\mu^2$ which includes $\mu=0$ and its neighbourhood (see Fig.~\ref{fig:tri}), we can anticipate different \blue{scaling laws} from the previous case where $a= O(\mu^2)$. The former region can be accessed by assuming $a\sim \varepsilon$ and letting either $\mu \sim \varepsilon$ or $\mu \ll \varepsilon$ such that $a\gg \mu^2$. In general, we can write, in the first approximation, $a=4\varepsilon + \mu>0 $ that includes all possible cases. It is easy to establish using~\eqref{disptri} that when $a\sim \varepsilon$, the fourth-order becomes negligible in comparison with the other two terms. Therefore, we may write
\begin{equation}
    \sigma = (4\varepsilon +\mu) k^2 - 4k^6. \label{triB}
\end{equation}

\section{Evolution of the flame front in the weakly nonlinear limit}

Let the equation of the perturbed flame front be $y=f(x,t)$ in the $x$-$y$ plane. In the linear theory, arbitrary $f$ can be constructed from the normal modes $e^{ikx+\sigma t}$ and the evolution of $f(x,t)$ for positive $\sigma$ is dictated by the function $\sigma(k)$ (equation~\eqref{triA} or~\eqref{triB}). However, in the absence of any instability, the quasi-planar front evolves in time $t$ according to
\begin{equation}
    f_t = - \frac{1}{2}f_x^2 + \frac{1}{8} f_x^4 + \cdots \label{geqn}
\end{equation}
which follows from the eikonal equation for flame-front propagation~\cite{buckmaster1983lectures}. Thus, in the presence of instability, nonlinear effects will become significant only when the perturbation amplitude grows to a magnitude such that $f_t \sim f_x^2$. In the weakly nonlinear limit, only the first-term on the right-hand side is retained. Such an analysis near type-II$_s$ bifurcation was first provided by Sivashinsky~\cite{sivashinsky1977nonlinear,sivashinsky1980flame}, who derived the Kuramoto--Sivashinsky (KS) equation for $f(x,t)$, describing the evolution of an unstable flame front. A similar analysis near type-I$_s$ bifurcation can be shown to yield the Swift-Hohenberg (SH) equation~\cite{cross1993pattern} for $f(x,t)$. Our attention here focuses on the neighbourhood of the tricritical point where neither of these equations is applicable.

\subsection{Equation for the critical regimes}

From~\eqref{triA}, it follows at once that the characteristic wavenumber is $k\sim \sqrt{|\mu|}$ and  the characteristic growth rate is $\sigma\sim |\mu|^3$. Inspection of equation~\eqref{geqn} using these length and time scales shows that the nonlinear term becomes important when $f\sim \mu^2$. Further by introducing the scalings 
\begin{equation}
    \tau =\frac{27}{16} |\mu|^3 t, \qquad \xi = \frac{\sqrt 3}{2}|\mu|^{1/2}x, \qquad F = \frac{4f}{9\mu^2}, \qquad q = \frac{4s}{9}
\end{equation}
the evolution equations for the flame front that account for both~\eqref{triA} and~\eqref{geqn} can be written as
\begin{align}
    &\text{Critical regime (type-II$_s$ side)}: &&  F_\tau + q F_{\xi\xi} + F_{\xi\xi\xi\xi}  -  F_{\xi\xi\xi\xi\xi\xi} + \frac{1}{2} F_\xi^2=0, \quad q>0, \label{criteq1}\\
    &\text{Critical regime (type-I$_s$ side)}: &&  F_\tau + q F_{\xi\xi} -  F_{\xi\xi\xi\xi}  -  F_{\xi\xi\xi\xi\xi\xi} + \frac{1}{2} F_\xi^2=0, \quad q>-\frac{1}{4}. \label{criteq2}
\end{align}
These two equations pertain, respectively, to the left and the right critical regimes, indicated in Fig.~\ref{fig:tri}, whereas the range of allowed values for $q$ in each case strictly corresponds to unstable states and excludes stable states.

\subsection{Equation for the tricritical regime}

Equation~\eqref{triB} suggests that the characteristic wavenumber in this case is $k\sim \varepsilon^{1/4}$ and the characteristic growth rate is $\sigma\sim \varepsilon^{3/2}$. In a similar manner as we did before, we identify that for the nonlinear effects to become important we need $f\sim \varepsilon$. Introducing the following rescalings
\begin{equation}
    \tau = (4\varepsilon+\mu)^{3/2}  t/2, \qquad \xi = (4\varepsilon+\mu)^{1/4}x/\sqrt 2, \qquad F=f/(4\varepsilon+\mu)
\end{equation}
we obtain the following governing equation in the weakly nonlinear limit
\begin{equation}
    F_\tau +  F_{\xi\xi} -  F_{\xi\xi\xi\xi\xi\xi} + \frac{1}{2} F_\xi^2=0. \label{trieq}
\end{equation}

\subsection{\blue{Sample numerical results of the sixth-order equations}}
Equations~\eqref{criteq1},~\eqref{criteq2} and~\eqref{trieq} are typically solved in a periodic domain with a period $2\pi L$. Integration of these equations \blue{over} this domain shows that the mean value \blue{$\frac{1}{2\pi L}\int_0^{2\pi L} F\,d\xi$} has a negative drift in time. To avoid this drift, one usually works with the variable $G=F_\xi$ in place of $F$, for which the mean value \blue{$\frac{1}{2\pi L}\int_0^{2\pi L} G\,d\xi$} is a constant. Measuring $\xi$ in units of $L$ and $\tau$ in units of $L^2$, we can simplify~\eqref{trieq} (using the same symbols for the rescaled $\xi$ and $\tau$) to
\begin{equation}
    G_\tau - \nu G_{\xi\xi\xi\xi\xi\xi}  + G_{\xi\xi}+GG_{\xi} =0,  \label{Gtri}
\end{equation}
where $\nu=1/L^4$. Under the same scaling, equations~\eqref{criteq1} and~\eqref{criteq2} simplify to
\begin{equation}
    G_\tau - \nu G_{\xi\xi\xi\xi\xi\xi} \pm\sqrt{\nu}G_{\xi\xi\xi\xi}  + qG_{\xi\xi}+GG_{\xi}  =0.  \label{Gcrit}
\end{equation}
All these equations can be integrated using \blue{periodic boundary} conditions and an initial condition, given by
\begin{equation}
    G(\xi,\tau) = G(\xi+2\pi, \tau), \qquad G(\xi,0) = G_0(\xi). \label{boun}
\end{equation}
\blue{For the sake of comparison, we shall also solve the fourth-order Kuramoto--Sivashinsky equation which, under suitable scaling, can be written in the form}
\begin{equation}
    \blue{G_\tau + \sqrt{\nu}G_{\xi\xi\xi\xi} + G_{\xi\xi} + GG_{\xi} =0}, \label{KS}
\end{equation}
\blue{subject to the conditions~\eqref{boun}. A useful quantity  to track as a function of time  in the computations is the gradient energy of the flame front which is defined by}
\begin{equation}
   \blue{E(\tau)=\frac{1}{2}\int_0^{2\pi}G^2\, d\xi=\frac{1}{2}\int_0^{2\pi} F_\xi^2\,d\xi.}
\end{equation}

\begin{figure}[h!]
\centering
\advance\leftskip-0.5cm
\includegraphics[scale=0.6]{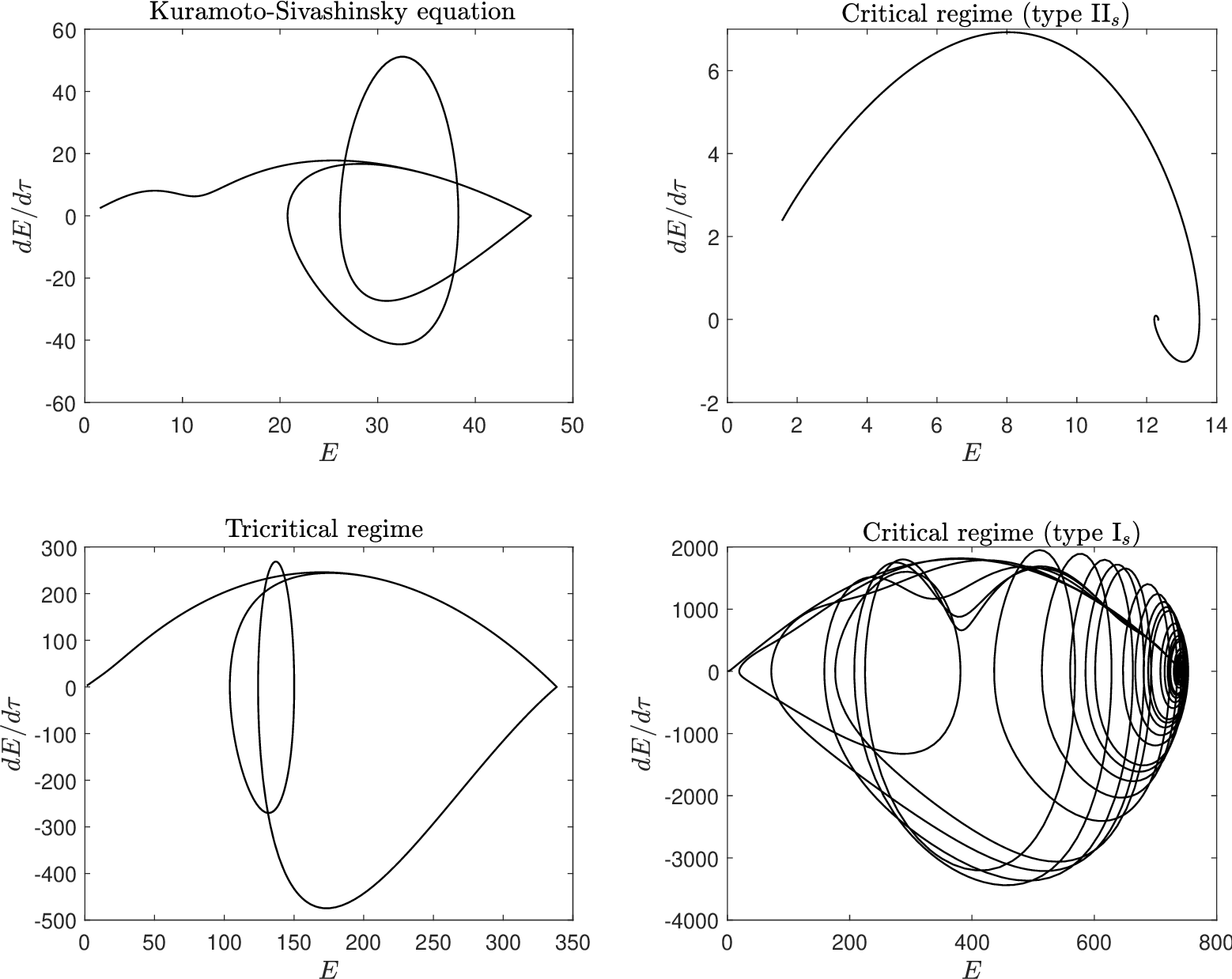}
\caption{\blue{Phase portrait of the function $E(\tau)$ for $\nu=0.04$ for the Kuramoto--Sivashinsky equation~\eqref{KS} and for the three equations~\eqref{Gtri}-\eqref{Gcrit} (with $q=1$) applicable in the three regimes. The phase trajectories in these plot correspond to the time interval $\tau\in[0,60]$. For the selected case, the KS equation and the tricritical equation results in a homoclinic orbit (cf. figure 2 of~\cite{smyrlis1996computational}) in which the cusps seen in the figures are equilibrium points that connect a stable and an unstable manifold. In the critical regime (type-II$_s$), the solution approaches a steady state via a stable spiral. The critical regime (type-I$_s$) results in a Shilnikov-type orbit (cf. figure 4 of~\cite{jolly1990approximate}) in which the point $(E,dE/d/\tau)=(744,0)$ acts as a connector between a stable saddle and an unstable center.}} 
\label{fig:triphase}
\end{figure}

\begin{figure}[h!]
\centering
\advance\leftskip-0.5cm
\includegraphics[scale=0.6]{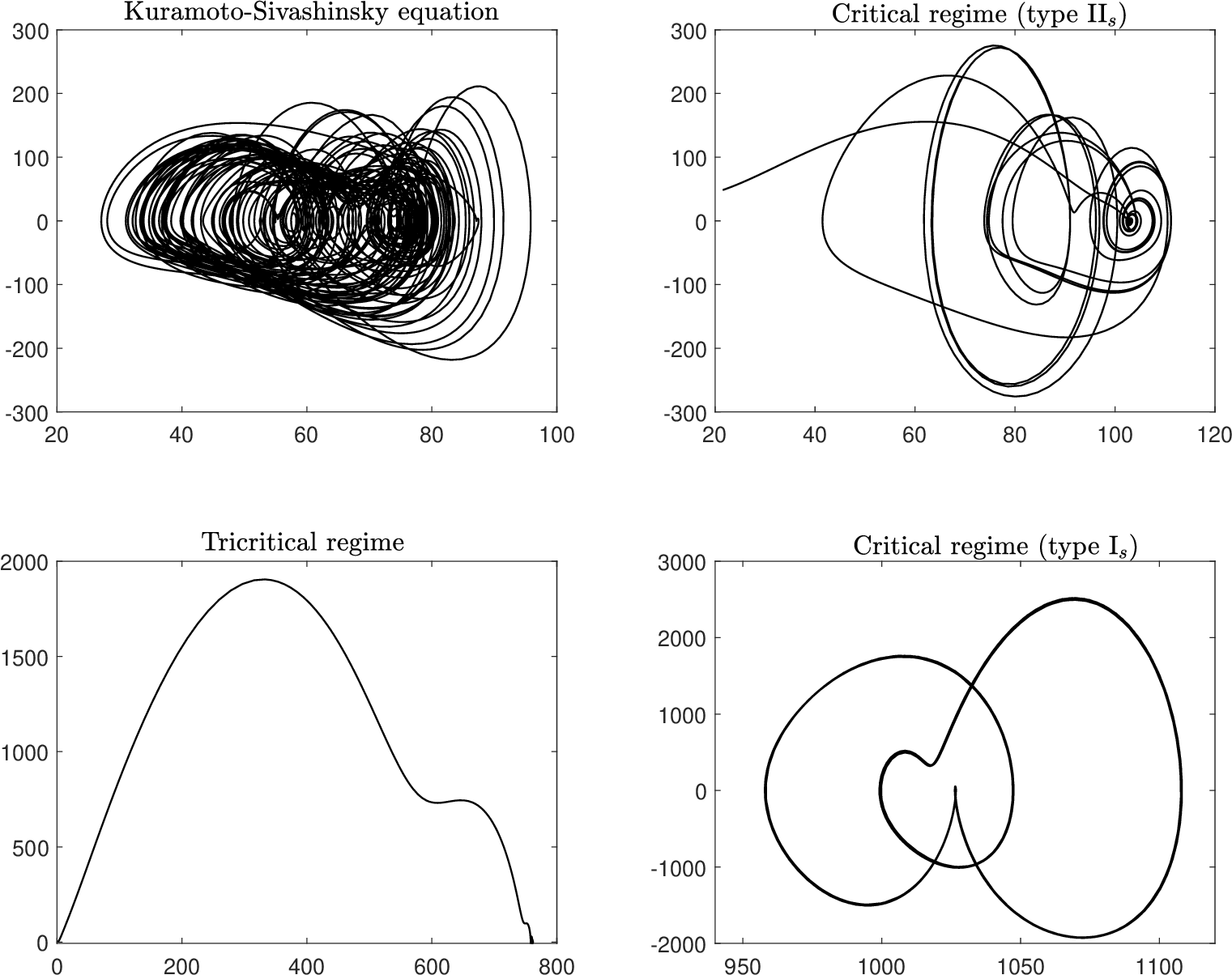}
\caption{\blue{Phase portrait of the function $E(\tau)$ for $\nu=0.005$ for the Kuramoto--Sivashinsky equation~\eqref{KS} and for the three equations~\eqref{Gtri}-\eqref{Gcrit} (with $q=1$) applicable in the three regimes. The phase trajectories in the first and the last plots correspond to the time intervals $\tau\in[41,68]$, whereas the time interval of the other two plots is $\tau\in[0,100]$. For this selected case, the KS equation displays a  chaotic oscillatory behaviour, whereas the tricritical equation has a stable steady solution at large times. In the critical regime  (type-II$_s$), the phase trajectory resembles a Shilnikov-type orbit mentioned in Fig.~\eqref{fig:triphase}. In the critical regime (type-I$_s$), we observe a periodic doubling event for the selected parametric values.}} 
\label{fig:triphase2}
\end{figure}

We have performed numerical computations using the initial condition $G_0=\cos\xi$ for selected values of $\nu$. Generally speaking, our computations suggest that the solutions of the equations~\eqref{Gtri}-~\eqref{boun} are qualitatively similar to that of the Kuramoto--Sivashinsky equation~\eqref{KS} with~\eqref{boun}, in that solutions approach non-trivial states only when $\nu<1$ and that there exist windows of $\nu$ with specific attractors~\cite{jolly1990approximate,smyrlis1991predicting,smyrlis1996computational}. The study of these attractors \blue{based on an exhaustive set of computations} and their connection to infinite-dimensional dynamical systems are of independent interest and deserve investigation in the future. \blue{For illustrative purposes, we present here the phase portraits of $E(\tau)$ computed from~\eqref{Gtri}-\eqref{KS} in Fig.~\ref{fig:triphase} for $\nu=0.04$ and in Fig.~\ref{fig:triphase2} for $\nu=0.005$. The description of these portraits are provided in the figure caption.}


\subsection{Generalized forms}
Since the type-II$_s$ bifurcation along with the nonlinear gradient term $(-f_x^2/2)$ appears in a number of other systems as well such as in slowly varying phase fields~\cite{kuramoto1978diffusion}, liquid interfacial problems~\cite{homsy1974model,nepomnyashchii1974stability,sivashinsky1980irregular,papageorgiou1990nonlinear}, trapped-ion instabilities~\cite{laquey1975nonlinear}, etc., it is plausible to expect a tricritical point (in the sense defined here) in these systems due to the variation of a new controlling parameter. In concluding this note, we thus write down the generalized three-dimensional form of the equations applicable in the neighbourhood of the tricritical point as
\begin{align}
    \text{Tricritical regime:} &\quad F_\tau + \nabla^2 F  - \nabla^6 F + \frac{1}{2}|\nabla F|^2 = 0,\\
    \text{Critical regimes:} &\quad F_\tau + q\nabla^2 F \pm \nabla^4 F - \nabla^6 F + \frac{1}{2}|\nabla F|^2 = 0,   
\end{align}
where $q>0$ when the plus sign applies and $q>-1/4$ when the minus sign applies. It is also apparent that the change of sign of the fourth-derivative term from one-side of the tricritical point to the other side, is mediated through the tricritical-regime equation in which the fourth-derivative term is absent.

\section{Summary}
The characteristics of cellular pattern in thick premixed flames, propagating in a direction transverse to the shear flow in a narrow channel, depend on the Lewis number and the Peclet number, as discussed in greater detail in~\cite{daou2023diffusive}. Consequently, the flame dynamics is controlled by the parameters $\mu$ and $\varepsilon$, which measure the extent of departure from the onset of instability in terms of the Lewis number and the Peclet number, respectively. Near the onset of cellular instability, the following scaling prevails:
\begin{align}
    & k\sim \sqrt{\mu},\quad \sigma \sim \mu^2,\quad f \sim \mu \quad \text{with}\quad  \varepsilon \sim \mu \quad \text{in} \quad \text{type II$_s$ bifurcation}, \label{ty2}\\
     & k-k_m\sim \sqrt{\mu},\quad \sigma \sim \mu,\quad f \sim \sqrt\mu \quad \text{with}\quad  \varepsilon \sim \mu \quad \text{in} \quad \text{type I$_s$ bifurcation} \label{ty1}
\end{align}
which respectively characterises the length scale (measured in units of laminar flame thickness), the time scale (measured in units of flame residence time) and the perturbation amplitude (measured in units of laminar flame thickness). Conventionally, only type II$_s$ bifurcation is encountered in the absence of the shear flow. In type I$_s$ bifurcation, the cells are not longwave type because $k_m\sim O(1)$ is not a small number.

On the other hand, in the three regimes in the neighbourhood of the tricritical point, we have
\begin{align}
    & k\sim \sqrt{\mu},\quad \sigma \sim \mu^3,\quad f \sim \mu^2 \quad \text{with}\quad  4\varepsilon+\mu \sim \mu^2 \quad \text{in the} \quad \text{critical regimes}, \label{cri}\\
     & k\sim \varepsilon^{1/4},\quad \sigma \sim \varepsilon^{3/2},\quad f \sim \varepsilon \quad \text{with}\quad  4\varepsilon+\mu \sim \varepsilon \quad \text{in the} \quad \text{tricritical regime.} \label{tri}
\end{align}
The key controlling small parameter is $\mu$ in the critical regimes and $\varepsilon$ in the tricritical regime, unlike in~\eqref{ty2}-\eqref{ty1} where one is allowed to arbitrarily select $\mu$ or $\varepsilon$ as the small parameter.  

\section*{Acknowledgements}
This research was supported by the UK EPSRC through grant EP/V004840/1.

\bibliographystyle{elsarticle-num}

\bibliography{sample}

\end{document}